\documentclass[aps,amsmath,amssymb,prb,twocolumn]{revtex4}   
\usepackage{graphicx}

\newcommand{\be}{\begin{equation}}
\newcommand{\ee}{\end{equation}}
\newcommand{\nn}{\nonumber}
\newcommand{\ef}{\mathrm{eff}}
\newcommand{\m}{\mathrm{max}}

\begin{document}

\title{A computer controlled pendulum with position readout}
\author{H. Hauptfleisch$^2$, T. Gasenzer$^1$, K. Meier$^2$, O. Nachtmann$^1$, and J. Schemmel$^2$}
  \affiliation{$^1$Institut f\"ur Theoretische Physik, Universit\"at Heidelberg, Philosophenweg 16, 69120 Heidelberg, Germany\\
  $^2$Kirchhoff-Institut f\"ur Physik, Universit\"at Heidelberg, INF 227, 69120 Heidelberg, Germany}

\date{\today}

\begin{abstract}
We have designed, built and operated a physical pendulum which allows one to demonstrate experimentally the behaviour of the pendulum under any equation of motion for such a device for any initial conditions. All parameters in the equation of motion can be defined by the user. The potential of the apparatus reaches from demonstrating simple undamped harmonic oscillations to complex chaotic behaviour of the pendulum. The position data of the pendulum as well as derived kinematical quantities like velocity and acceleration can be stored for later offline analysis.
\end{abstract}

\maketitle

\section{Introduction}
The equation of motion for a simple damped and driven pendulum is easy to find, yet its solutions are phenomenologically rich. 
By changing the parameters, the behaviour of the pendulum may change from periodic to chaotic. 
To demonstrate all possible phenomena with a single device it is necessary to apply a well-defined and time-dependent external torque at any given angular position of the pendulum.

The idea presented in this paper is to connect the pendulum both with an angular transducer, which sends angular position data of the pendulum to a computer, and with a torsion spring connected to a stepping motor. 
The angle $\psi$ between the rotational positions of the axes of the stepping motor and the pendulum, respectively, determines a torque that may depend on the position of the pendulum, its angular velocity, and on time. The stepping motor is controlled in a closed control loop formed by the angular transducer, a control computer, and the motor itself.

Apart from being used in the described closed control loop, the angular transducer also delivers angular position data, with a precision of 8192 steps for one full turn of 360 degrees. This data-recording feature allows for detailed comparisons with computer simulations \cite{Baker1995a,Shew1999a,Blackburn1992a,Blackburn1992b,Baker1996a}.

This paper is organised as follows. Section \ref{aufbau} describes the mechanical and electronics set up. Software concepts are discussed in Sect.~\ref{software}. In Sections \ref{messung} and \ref{results} we collect and discuss several results of measurements demonstrating a broad range of applications and the usefulness of the device for teaching the many different aspects of the solutions of the equation of motion for a physical pendulum.

\section{The apparatus}

\subsection{Mechanical Set up and Electronics}
\label{aufbau}
\begin{figure}[tb]
\begin{center}
\includegraphics[width=\columnwidth]{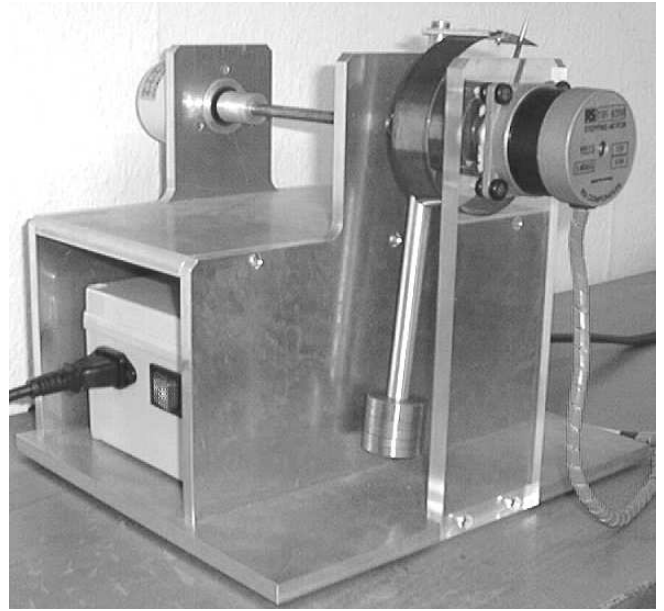}
\caption{\small 
Photograph of the pendulum.}
\label{pfoto}
\end{center}
\end{figure}

\begin{figure}[tb]
\begin{center}
\includegraphics[width=0.7\columnwidth]{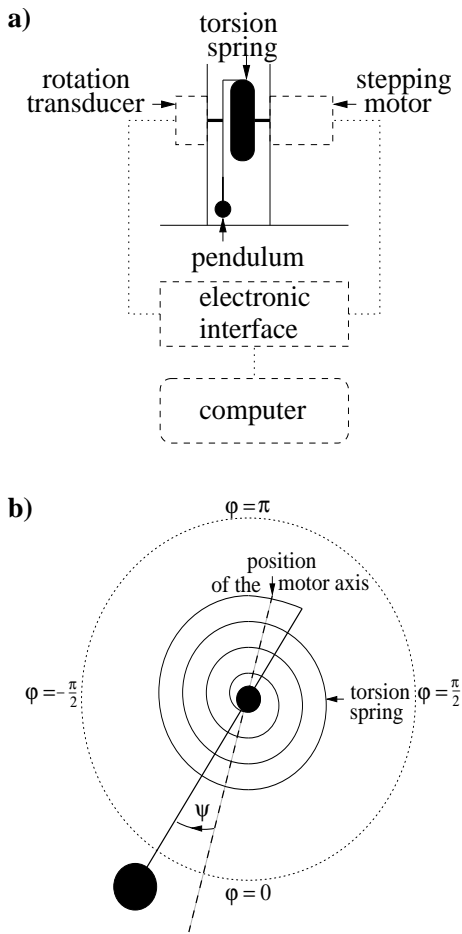}
\caption{\small 
Panel (a) shows the mechanical setup of the pendulum,
(b) the definition of the absolute position $\varphi$ of the pendulum and of the angle $\psi$ between the axes of the pendulum and the stepping motor.}
\label{paufbau}
\end{center}
\end{figure}

The mechanical set up consists of an aluminium base structure holding the pendulum, the angular transducer, the stepping motor, and the torsion spring. 
Figure \ref{pfoto} shows a photograph of the pendulum; a schematic representation is shown in Fig. \ref{paufbau}(a).

The body of the pendulum is made from stainless steel and consists of a rod with a length of $0.20\,$m and a set of steel disks to adjust its mass and moment of inertia. 
By changing the number of steel disks the total mass of the moving part can be adjusted from $0.110\,$kg to $0.288\,$kg and the corresponding moment of inertia varies between $1.56\times10^{-3}\,$kg m$^2$ and $6.61\times10^{-3}\,$kg m$^2$. 
The steel suspension axis divides the pendulum rod in a length ratio of one to three. It is mounted with ball bearings in the aluminium base structure. 
The axis is connected to the angular transducer. 
The transducer works optically with a minimum of internal friction. 
It provides a precision of 8192 steps for a full turn of 360 degrees corresponding to $0.77\,$mrad per step.  
The resulting internal friction of the pendulum is caused mainly by the angular transducer and by imperfections of the ball bearings holding the steel axis in the base structure. 
The position data of the transducer are sent to a digital I/O-card 
(DAQCard$^\mathrm{TM}$-DIO-24, National Instruments) placed in the PCI port 
of the control computer. 
A program converts the grey-code of the optical angle encoder into a numerical value for the angle $\varphi(t)$, representing the actual angular position of the pendulum at time $t$, as shown in Fig.~\ref{paufbau}(b). 
The angular position has a periodicity of $2\pi$. 

The stepping motor allows driving and holding the motor axis at a well-defined angular position. 
The employed hybrid motor divides one full rotation into 400 digital steps corresponding to $15.71\,$mrad per step. 
The signals for controlling the motor are generated on a drive board connected to the digital I/O-card of the control computer.  
The torsion spring is made of a thin steel band and was manufactured by a watch-maker.
Its spring constant has been determined to $0.251\,$Nm/rad.

\subsection{Software Concept}
\label{software}
\subsubsection{The equation of motion}
According to Newton's second law, the sum of all acting torques is equal to the product of the moment of inertia $I$ and the angular acceleration  $\ddot\varphi$ of the pendulum:
\begin{equation}
  I \ddot \varphi 
  =  T^{*} \left( \varphi, \dot \varphi, t \right)
      -T_{r,\mathrm{max}} \sin\varphi
      -b\, \dot\varphi 
      -c~\mathrm{sgn}(\dot\varphi). 
\label{eq:eom}
\end{equation}
The sum of all acting torques receives contributions from the torque $T^*$ generated by the torsion spring via the stepping motor, the restoring torque $-T_{r,\mathrm{max}}\sin \varphi$ of the gravitational field, a linear friction term $-b~\dot\varphi$ (caused by the air and the control system) and a constant friction term $-c~\mathrm{sgn}(\dot\varphi)$ (caused by the internal friction of the ball bearings and the rotation transducer). 
The value of the function $\mathrm{sgn}(\dot\varphi)$ is defined as $1$ for $\dot\varphi>0$, as $-1$ for $\dot\varphi<0$, and as $0$ for $\dot\varphi=0$.

The user is able to vary the torque by controlling the stepping motor. 
The computer control of the motor has been constructed such that the torque generated by the motor has the following functional form
\begin{equation}
  T^{*}\left( \varphi, \dot \varphi, t \right) 
  =  T_{0}^{*} \sin\left(\omega t \right) 
      +T_{r,\mathrm{max}}^{*} \sin\varphi 
      +b^{*}\,\dot\varphi 
      +c^*~\mathrm{sgn}(\dot\varphi).
\label{eq:usertorque}
\end{equation}
Here the angular frequency $\omega$ and the set of four parameters, $T^*_0,T^*_{r,\mathrm{max}},b^*$ and $c^*$ are adjustable by the user. 
Inserting the parametrisation of the torque $T^*$, Eq.~(\ref{eq:usertorque}), into Eq.~(\ref{eq:eom}) yields the equation of motion 
\begin{eqnarray}
\label{eq:eomuser}
  I \ddot \varphi 
  =-T_{r,\mathrm{eff}}\sin\varphi+T^*_0\sin(\omega t)
-b_{\mathrm{eff}}\dot\varphi-c_{\mathrm{eff}}\mathrm{sgn}(\dot\varphi)
\end{eqnarray}
where
\begin{align}
\label{3a}
  T_{r,\ef}
  &=T_{r,\m}-T^*_{r,\m},
  \nn\\
  b_{\ef}
  &=b-b^*,
  \nn\\
  c_{\ef}
  &=c-c^*.
\end{align}
The physical meaning of the four terms on the r.h.s. of Eq.~(\ref{eq:eomuser}) is as follows. The first term represents the effective gravitational acceleration, the second term is the harmonic driving term, and the last two terms represent the effective friction. The linear or viscous friction term $\propto\dot\varphi$ is governed by the parameter $b_{\ef}$ the (piecewise) constant friction term $\propto\mathrm{sgn}(\dot\varphi)$ by the parameter $c_{\ef}$. Changing the parameters marked with * the software control of the stepping motor allows one to set the effective parameters for gravitational acceleration, driving and damping to any value required. 
This key feature of the system even allows one to compensate for the friction terms completely or to set the effective acceleration of the pendulum in the earth's gravitational field to any value including zero. In the following we shall, however, always suppose that the effective friction parameters satisfy
\begin{align}
\label{3b}
  b_{\ef}
  &\geq 0,
  \nn\\
  c_{\ef}
  &\geq 0.
\end{align}
%

\subsubsection{The control system}
The software which has been written to achieve this flexibility basically consists of a closed control loop, which sets the correct position of the stepping motor at any position and velocity of the pendulum at any time. 
A graphical user interface (GUI) allows convenient control of the parameters of the equation of motion, Eq.~(\ref{eq:eomuser}), for setting up the experiment before or even during motion.

The angular distortion $\psi$ of the spring depends linearly on the torque $T^*$,
\begin{equation}
   \psi\left( \varphi, \dot \varphi, t \right)
   = -\frac{T^{*}\left( \varphi, \dot\varphi, t \right)}{D_{0}},
\label{eq:hooke}
\end{equation}
with the inverse of the spring constant $D_0$ as proportionality factor.  
This relation defines the angle $\psi$ between the axes of the stepping motor and the pendulum which is required to obtain the desired torque. 
        
The actual position of the stepping motor is determined by the closed control loop. 
In each run through the loop the actual positions of the pendulum and the stepping motor are compared and, if necessary, the stepping motor is moved one step towards the required position. 
The speed of the closed control loop is a linear function of the last measured deviation between the actual and the required position of the stepping motor as long as the maximum stepping rate of the stepping motor is not exceeded. The maximum stepping rate is $1300\,$Hz. 
As indicated in Eq.~(\ref{eq:usertorque}) the closed control system also induces linear and constant friction terms. Their strengths depend on the parameters $b^*$ and $c^*$, respectively. Through the closed control loop these parameters can be adjusted by the user via the GUI. 
A special method of measurement, which is discussed in Sect.~\ref{messung}, allows one to find out the actual values of these constants $b^*$ and $c^*$. The angular velocity of the stepping motor is determined by the angular velocity of the pendulum $\dot\varphi$ and the corresponding angular velocity of the motor axis as required to produce the correct angle $\psi$ for the desired time-dependent torque $T^{*}( \varphi, \dot \varphi, t)$, related to $\psi$ by Eq.~(\ref{eq:hooke}). 
The parameters have to be chosen such that the maximum stepping rate of the stepping motor is not exceeded. 
A critical parameter for the set up of an experiment is the limited maximum holding torque of the stepping motor ($0.494\,$Nm) to which the mass and the maximum acceleration of the pendulum have to be adjusted. 
The graphical user-interface has been programmed in C++ using the QT- and Qwt-libraries for the generation of graphical elements. 
The parameters of the closed control loop, as well as the parameters of the equation of motion, can be selected with the help of various sliders. 
Several options for measurements can be selected by the user. 
All recorded position data are saved and can be plotted in different representations.

\section{Results for a Non-Driven Pendulum}
\label{messung}

Without external driving torque, $T^*_{0}=0$, the oscillation of the pendulum exhibits specific different characteristics depending on the selected damping terms. 

\subsection{Small elongation and linear friction}
\label{smallelonglinfric}

\begin{figure}[tb]
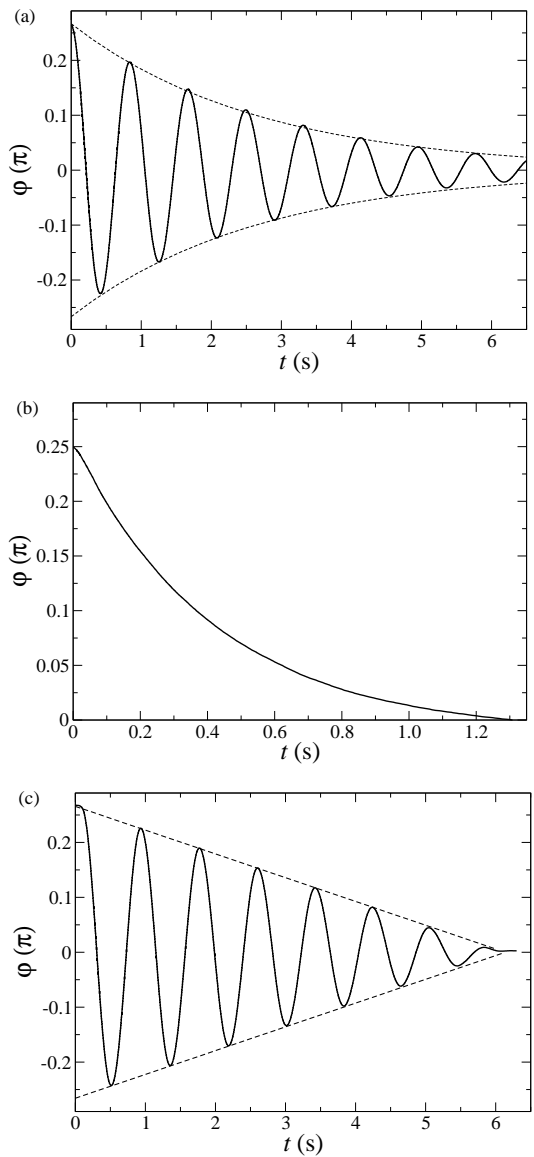

\begin{center}
\includegraphics[width=0.8\columnwidth]{fig3a.eps}
\vspace*{3mm}\\
\includegraphics[width=0.8\columnwidth]{fig3b.eps}
\vspace*{3mm}\\
\includegraphics[width=0.8\columnwidth]{fig3c.eps}
\caption{\small 
Different forms of motion of the non-driven pendulum at small elongations. Panels (a) and (b) show results where only the linear friction term is present, see Eq.~(\ref{4a}). Panel (a) realizes the underdamped, panel (b) the overdamped case, see Eqs.~(\ref{8}) and (\ref{11}), respectively. Panel (c) shows the case of small elongations with a piecewise constant friction term, see Eq.~(\ref{30}). There the amplitude of the oscillation decreases linearly with time.}
\label{reibung}
\end{center}
\end{figure}

\noindent
In this subsection we suppose that the parameters of the pendulum are chosen such that
\begin{align}
\label{4a}
  T_{r,\ef}
  &>0,
  \nn\\
  b_{\ef}
  &>0,
  \nn\\
  c_{\ef}
  &=0.
\end{align}
We also suppose that the amplitude of the oscillation is much smaller than $\pi/2$. Then the nonlinear equation of motion, Eq.~(\ref{eq:eomuser}), can be approximated by using $\sin\varphi\sim\varphi$ which leads to
\be
\label{5}
  \ddot\varphi+\frac{2}{\tau}\dot\varphi+\omega^2_0\varphi
  =0.
\ee
Here we define
\begin{align}
\label{6}
  \tau
  &=\frac{2I}{b_{\ef}},
  \nn\\
  \omega^2_0
  &=\frac{T_{r,\ef}}{I}.
\end{align}
The solutions of Eq.~(\ref{5}) are discussed in many textbooks \cite{6}. We have to distinguish three cases. 
\begin{itemize}
\item[(i)] Underdamping (oscillatory motion)

This happens for
\be
\label{7}
  \omega^2_0>\frac{1}{\tau^2}.
\ee
The general solution reads for this case
\be
\label{8}
  \varphi(t)
  =\varphi_0\exp(- t/\tau)\cos(\bar\omega t+\bar\delta)
\ee
where
\be
\label{9}
  \bar\omega
  =\sqrt{\omega^2_0-\left(\frac1\tau\right)^2}.
\ee
The free parameters of the solution, Eq.~(\ref{8}), are $\varphi_0$ and $\bar\delta$. The realization of such a pendulum motion is shown in Fig.~\ref{reibung}(a).
\item[(ii)] Overdamping

\noindent
This happens for 
\be
\label{10}
  \omega^2_0
  <\frac{1}{\tau^2}.
\ee
Here the general solution of Eq.~(\ref{5}) reads
\begin{align}
\label{11}
  \varphi(t)
  &=\varphi_+
    \exp \left[-\left(\frac1\tau+\sqrt{\frac{1}{\tau^2}-\omega^2_0}\right)t\right]
  \nn\\
  &+\ \varphi_-\exp\left[-\left(\frac1\tau-
      \sqrt{\frac{1}{\tau^2}-\omega^2_0}\right)t\right]
\end{align}
with $\varphi_+$ and $\varphi_-$ free constants. An experimental realization of this case with our apparatus is shown in Fig.~\ref{reibung}(b).
\item[(iii)] Critical damping

\noindent
This happens for 
\be
\label{12}
   \omega^2_0
   =\frac{1}{\tau^2}.
\ee
The general solution of Eq.~(\ref{5})  for this case reads
\be
\label{13}
   \varphi(t)
   =\left(\varphi_1+\varphi_2\frac{t}{\tau}\right)\exp
   \left(-\frac t\tau\right)
\ee
where $\varphi_1$ and $\varphi_2$ are free constants. 
\end{itemize}

\subsection{Small elongation and piecewise constant friction}
\label{smallelongconstfric}

\noindent
Here we suppose that the parameters of the pendulum have been chosen such that
\begin{align}
\label{14}
   T_{r,\ef}
   &>0,
   \nn\\
   b_{\ef}
   &=0,
   \nn\\
   c_{\ef}
   &>0.
\end{align}
For small oscillations we thus obtain, from Eq.~(\ref{eq:eomuser}), the equation of motion 
\be
\label{15}
  \ddot\varphi(t)+\omega^2_0~\varphi(t)+\omega^2_0\chi_0\,
  \mathrm{sgn}\big(\dot\varphi(t)\big)
  =0
\ee
where $\omega^2_0$ is as in Eq.~(\ref{6}) and 
\be
\label{16}
  \chi_0
  =\frac{c_{\ef}}{\omega^2_0I}.
\ee
For initial conditions which allow for an oscillatory solution of Eq.~(\ref{15}) this solutions can be constructed in the following way. 
Writing the elongation and its time derivative at the initial time $t_{i}$ as  
\begin{align}
\label{17}
  \varphi_i
  &=\varphi(t_i),
  \nn\\
  \dot\varphi_i
  &=\dot\varphi(t_i)
\end{align}
a solution of Eq.~(\ref{15}) is given by
\begin{align}
\label{18}
  \varphi(t)
   &=-\chi_0~\mathrm{sgn}(\dot\varphi_i)
   +\big[\varphi_i+\chi_0~\mathrm{sgn}(\dot\varphi_i)\big]
   \cos\big[\omega_0(t-t_i)\big]
   \nonumber\\
   &\quad
   +\ \frac{1}{\omega_0}\dot\varphi_i\sin\big[\omega_0(t-t_i)\big].
\end{align}
Inserting the second time derivative of this expression into Eq.~(\ref{15}) one finds that
\begin{align}
\label{18a}
  &\ddot\varphi(t)+\omega^2_0~\varphi(t)+\omega^2_0\chi_0\,
  \mathrm{sgn}\big(\dot\varphi(t)\big)
  \nonumber\\
  &\quad
  =\ \omega^2_0\chi_0\,[
   \mathrm{sgn}\big(\dot\varphi(t)\big)-\mathrm{sgn}\big(\dot\varphi_{i}\big)].
\end{align}
Hence, Eq.~(\ref{18}) solves the equation of motion (\ref{15}) only as long as the angular velocity does not change its sign.
We illustrate in the following how Eq.~(\ref{18}) can be used to iteratively construct a damped oscillatory solution of Eq.~(\ref{15}).

We choose initial conditions where for $t_i=0$ the pendulum starts at an elongation
\be
\label{19}
  \varphi_i
  =\varphi_0>\chi_0
\ee
with infinitesimally negative velocity
\be
\label{20}
  \dot\varphi_i=-\epsilon.
\ee
Here and in the following $\epsilon>0$ denotes an arbitrarily small positive number. For this choice Eq.~(\ref{18}) implies that
\begin{align}
\label{21}
  \varphi(t)
  &=\chi_0+(\varphi_0-\chi_0)\cos (\omega_0t),
  \nn\\
  \dot\varphi(t)
  &=-(\varphi_0-\chi_0)~\omega_0\sin(\omega_0t),
  \nn\\
  \mathrm{sgn}\big(\dot\varphi(t)\big)
  &=-1\quad\mathrm{for}\quad 0<t<\frac{\pi}{\omega_0}.
\end{align}
It is easy to check that Eq.~(\ref{21}) gives indeed a solution of Eq.~(\ref{15}) for 
\be
\label{22}
  0<t<\pi/\omega_0,
\ee
that is, until the pendulum reaches its first turning point.
If $\varphi_0>2\chi_0$ this turning point is located at the negative angle
\begin{align}
\label{23}
  \varphi\left(\frac{\pi}{\omega_0}\right)
  &=-\varphi^{(1)}_0,
  \nn\\
  \varphi^{(1)}_0
  &=\varphi_0-2\chi_0.
\end{align}

Now suppose that the initial elongation was $\varphi_0>3\chi_0$, i.e.,
\be
\label{24}
  \varphi^{(1)}_0>\chi_0.
\ee
In this case we can continue to construct the solution of Eq.~(\ref{15}), for a further half oscillation period, by setting in Eq.~(\ref{18})
\begin{align}
\label{25}
  t_i
  &=\pi/\omega_0,
  \nn\\
  \varphi_i
  &=\varphi\left(\frac{\pi}{\omega_0}\right)=-\varphi^{(1)}_0,
  \nn\\
  \dot\varphi^{(1)}_i
  &=\epsilon.
\end{align}
For times
\be
\label{27}
  \frac{\pi}{\omega_0}<t<\frac{2\pi}{\omega_0}
\ee
the solution is given by
\begin{align}
\label{26}
  \varphi(t)
  &=-\chi_0-(\varphi^{(1)}_0-\chi_0)\cos(\omega_0t-\pi),
  \nn\\
  \dot\varphi(t)
  &=(\varphi^{(1)}_0-\chi_0)~\omega_0\sin(\omega_0t-\pi),
  \nn\\
  \mathrm{sgn}\big(\dot\varphi(t)\big)
  &=+1.
\end{align}

Clearly, we can continue this game, provided that we had chosen $\varphi_0>5\chi_0$ initially. For the elongations at the turning-point times, $t=n\pi/\omega_0~(n=0,1,2,\dots)$, we obtain
\be
\label{28}
  \varphi\left(\frac{n\pi}{\omega_0}\right)
  =(-1)^n[\varphi_0-2n\chi_0].
\ee
That is, we find a linear decrease of the maximal elongations with $n$. The above way of constructing a solution of Eq.~(\ref{15}) ceases to work, though, once the absolute value of the elongation at a turning point is smaller than $\chi_0$. We define, therefore, $n_{\m}$ as the positive integer with the following property
\be
\label{29}
  n_{\m}-1
  <\frac{\varphi_0-\chi_0}{2\chi_0}\leq n_{\m}.
\ee
The solution of Eq.~(\ref{15}) with the initial condition (\ref{19}), (\ref{20}) then reads  
\begin{align}
\label{30}
  \varphi(t)
  &=(-1)^n\Big\{\chi_0
   +\big[\varphi_0-(2n+1)\chi_0\big]\cos (\omega_0t-n\pi)\Big\}
  \nn\\
  \mathrm{for}\
  & \frac{n\pi}{\omega_0}\leq t<\frac{(n+1)\pi}{\omega_0},
   ~n=0,1\dots(n_\m-1).
\end{align}
An experimental example of such a motion of the pendulum is shown in Fig.~\ref{reibung}(c) where the linear decrease of the maximal elongations, as given theoretically in Eq.~(\ref{28}), is clearly visible.

But what happens with the solution (\ref{30}) for $t>n_\m\pi/\omega_0$? 
If $\varphi(n_\m\pi/\omega_0)=0$, the equilibrium has been reached precisely at time $t=n_\m\pi/\omega_0$. It can be shown that in all other cases the solution of Eq.~(\ref{15}) given in Eq.~(\ref{30}) cannot be extended for $t>n_\m\pi/\omega_0$. 

This simply means that for very small amplitudes $\varphi$ and velocities $\dot\varphi$ the approximation of the friction term by a function $c_\ef\mathrm{sgn}(\dot\varphi)$ ceases to make sense. In reality, the discontinuous function $\mathrm{sgn}(\dot\varphi)$ must then be replaced by a continuous function, for instance one behaving as $\dot\varphi$ for $\dot\varphi\to 0$. For very small amplitudes and velocities the motion of the pendulum will be described by one of the cases discussed in Subsection \ref{smallelonglinfric}. That is, the pendulum will return smoothly to its equilibrium position. 

From data like those shown in Fig.~\ref{reibung} the numerical values of the effective friction parameters $b_\ef$ and $c_\ef$ in Eq.~(\ref{eq:eomuser}) can be determined. First, the constant friction term is varied until the amplitude approaches an exponential decrease. This allows to read off from the data $b_\ef/I$ and, since $I$ is known, $b_\ef$. Then the linear friction term is varied until one reaches the undamped oscillation case, that is, absence of the linear friction term. Then the constant friction term can be tuned and determined from the data. This method of measurement is supported by theory only as long as the amplitude of the oscillation is small enough. 
Nevertheless the results are quite satisfying. 

\subsection{Large elongation and no friction}
\label{largeelongnofriction}

\noindent
Now we turn to the case of zero effective friction terms which can be easily realized by tuning the torque $T^*$, Eq.~(\ref{eq:usertorque}), appropriately. We abandon the approximation of small elongations of the pendulum and consider the exact equation of motion, Eq.~(\ref{eq:eomuser}), with $T_{r,\ef}>0$, $T_{0}^*=0$, $b_{\ef}=0$ and $c_\ef=0$:
\be
\label{31}
  \ddot\varphi(t)+\omega^2_0\sin \varphi(t)=0
\ee
where again
\be
\label{31a}
  \omega^2_0=T_{r,\ef}/I.
\ee
To solve (\ref{31}) we take as usual\cite{6} the route over the energy theorem. This gives
\be
\label{32}
  \frac12\dot\varphi^2-\omega^2_0\cos \varphi
  =\eta=\mathrm{const}.
\ee
where, clearly, we must have 
\be
\label{33}
  \eta\geq -\omega^2_0.
\ee
We shall discuss here only the case
\be
\label{34}
  -\omega_0^2\leq \eta<\omega^2_0
\ee
and set
\begin{align}
\label{35}
  \eta
  &=-\omega^2_0\cos\varphi_0,
  \nn\\
  0
  &\leq\varphi_0<\pi.
\end{align}
Denoting again the initial conditions at $t=t_i$ as
\begin{align}
\label{36}
  \varphi(t_i)
  &=\varphi_i,
  \nn\\
  \dot\varphi(t_i)
  &=\dot\varphi_i
\end{align}
the solution of Eq.~(\ref{31}) in determined, via Eq.~(\ref{32}), by the integral equation
\begin{align}
\label{37}
  \int^{\varphi(t)}_{\varphi_i}
  \frac{d\varphi'}{\sqrt{\sin^2\frac{\varphi_0}{2}-\sin^2\frac{\varphi'}{2}}}
  =\mathrm{sgn}(\dot\varphi_i)~2\omega_0~(t-t_i).
\end{align}
Here 
\be
\label{38}
  \sin^2\frac{\varphi_0}{2}
  =\sin^2\frac{\varphi_i}{2}+\frac{1}{4\omega^2_0}
   (\dot\varphi_i)^2
\ee
and the initial conditions must be such that Eq.~(\ref{38}) has a solution with $0\leq\varphi_0<\pi$. The integral in Eq.~(\ref{37}) is of the elliptic type \cite{7}. The pendulum oscillates with a certain period $T$ between $\varphi_0$ and $-\varphi_0$. Contrary to the case of small oscillations, the period $T$ depends now on the amplitude $\varphi_0$. From Eq.~(\ref{37}) we find easily
\be
\label{39}
  T
  =\frac{2}{\omega_0}\int^{\varphi_0}_0
   \frac{d\varphi'}{\sqrt{\sin^2\frac{\varphi_0}{2}-\sin^2\frac{\varphi'}{2}}}.
\ee
This can be related to the complete elliptic integral of the first kind\cite{7}
\begin{align}
\label{40}
  T
  &=\frac{4}{\omega_0}K\left(\sin\frac{\varphi_0}{2}\right)
  \nn\\
  &=\frac{4}{\omega_0}\int^{\pi/2}_0
    \frac{d\alpha}{\sqrt{1-\sin^2\frac{\varphi_0}{2}\sin^2\alpha}}.
\end{align}
%
\begin{figure}[tb]
\begin{center}
\includegraphics[width=0.8\columnwidth]{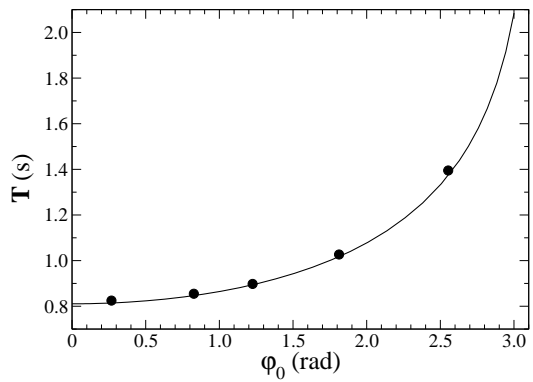}
\caption{\small 
Dependence of the period of oscillation $T$, Eq.~(\ref{40}), on the maximum elongation $\varphi_{0}$. The data points are compared to the calculation which is represented by the solid line.}
\label{schwinger}
\end{center}
\end{figure} 
Figure~\ref{schwinger} shows the calculated dependence of this oscillation period $T$ on the maximum elongation $\varphi_0$ and the verification of this curve by five measurements.

The period of oscillation also depends on the value of the effective gravitational restoring torque $T_{r,\ef}$, see Eq.~(\ref{31a}). This can be varied by changing the mass of the pendulum, which is quite easy to do, or the effective acceleration due to gravity, which seems to be impossible without leaving the earth or at least its surface. For our pendulum we can vary the effective gravitational torque $T_{r,\mathrm{eff}}$ by varying the parameter $T_{r,\m}^{*}$, see Eqs.~(\ref{eq:usertorque})--(\ref{3a}). At each position of the pendulum the torsion spring adds a well-defined torque. In this way it is possible to demonstrate an oscillation like that which would occur on the moon or without any gravitational field. In the latter case we have $\omega_0=0$ and the pendulum just moves with constant angular velocity. This case corresponds to the situation of a pendulum in empty space or in a freely falling laboratory in a gravitational field. That is, we can simulate the conditions prevailing in a space laboratory like the ISS.

\section{Results for a Driven Pendulum}
\label{results}

\subsection{Resonance phenomena}
\label{sec:regular}
Here we consider an external driving of the pendulum in a periodic fashion but with small enough amplitude of the driving torque such that the pendulum never reaches the top position. 
This leads in general, after the transients have disappeared, to an oscillation with frequencies related to that of the periodic driving torque. In this regime we can study resonance phenomena with our device.
Tuning the constant friction term in Eq.~(\ref{eq:eomuser}) to zero, $c_\ef=0$, and supposing $T_{r,\ef}>0$ and $b_\ef>0$ our pendulum's motion is governed by the equation 
\be
\label{41}
  \ddot\varphi +\frac2\tau \dot\varphi+\omega^2_0\sin\varphi
  =\frac{T^*_0}{I}\sin(\omega t).
\ee
Here $\omega^2_0$ and $\tau$ are as in Eq.~(\ref{6}). We shall discuss here only the theory for small elongations. Approximating $\sin \varphi$ by $\varphi$ in Eq.~(\ref{41}) we get
\be
\label{42}
  \ddot\varphi+\frac2\tau\dot\varphi+\omega^2_0\varphi
  =\frac{T^*_0}{I}\sin(\omega t).
\ee
Again, as in Sect.~\ref{smallelonglinfric}, we have to distinguish the underdamped, the overdamped, and the critically damped cases. As an example we give the well-known solution for the underdamped case
\begin{align}
\label{43}
  \varphi(t)
  &=\bar\varphi_0\exp(-t/\tau)\cos(\bar\omega t+\bar\delta)
  \nn\\
  &+\ \varphi_0\sin(\omega t+\delta)
\end{align}
where $\bar\omega$ is given by Eq.~(\ref{9}) and 
\begin{align}
\label{44}
  \varphi_0
  &=\frac{T^*_0}{I}
    \left[(\omega^2-\omega^2_0)^2+\frac{4\omega^2}{\tau^2}\right]^{-1/2},
  \nn\\
  \tan\delta
  &=-\frac{2\omega}{\tau(\omega^2_0-\omega^2)}.
\end{align}
%
\begin{figure}[tb]
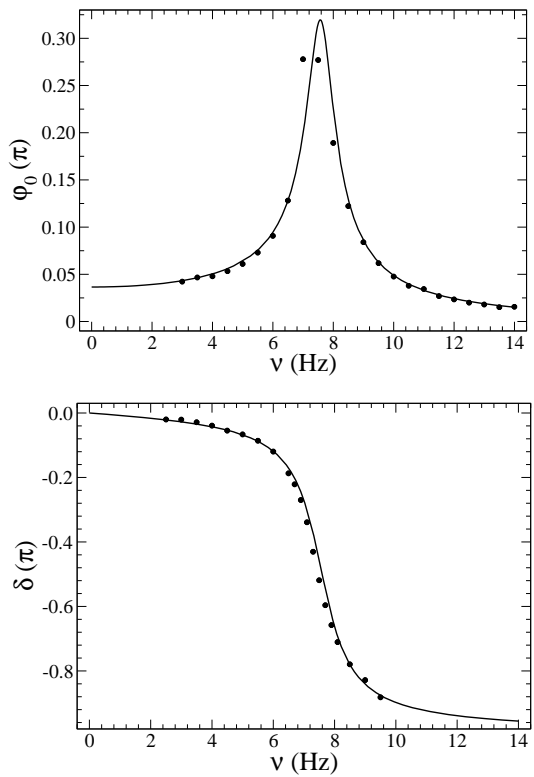

\begin{center}
\includegraphics[width=0.8\columnwidth]{fig5a.eps}
\vspace*{3mm}\\
\includegraphics[width=0.8\columnwidth]{fig5b.eps}
\caption{\small 
Amplitude $\varphi_0$ and phase shift $\delta$ of a pendulum driven by a periodic torque of the frequency $\nu$. The curves are the analytical results of the linearised equation of motion, see Eq.~(\ref{44}). The curves differ from the data points for the the frequency range of large amplitudes due to the non-linear terms in Eq.~(\ref{41}).}
\label{resonanz}
\end{center}
\end{figure} 
The constants $\bar\varphi_0$ and $\bar\delta$ are arbitrary and are needed to fulfil general initial conditions. After the transients, that is, the terms $\propto\exp(-t/\tau)$ in Eq.~(\ref{43}), have decayed the maximal elongation of the pendulum, $\varphi_0$, and the phase shift, $\delta$, can be measured as functions of the driving frequency $\nu=\omega/(2\pi)$. The measured points shown in Fig.~\ref{resonanz} are compared to the results of the calculation for small elongations in Eq.~(\ref{44}). The data points in Fig.~\ref{resonanz} follow nicely the theoretical curves except at the peak of the amplitude in the resonance region, that is for $\nu\approx 7\,$Hz. There the amplitude $\varphi_0$ gets large and the approximation $\sin\varphi\sim\varphi$ becomes bad. 

\begin{figure}[tb]
\begin{center}
\includegraphics[width=0.8\columnwidth]{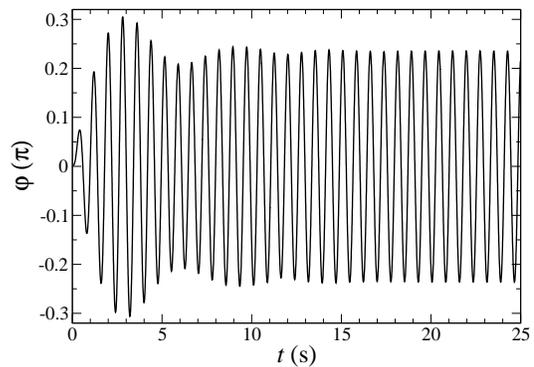}
\caption{\small 
Elongation $\varphi(t)$ of the externally driven pendulum as function of the time $t$. For time $0\leq t\leq 15\,$s the effect of the transients is clearly visible.}
\label{einschwing}
\end{center}
\end{figure}
In Fig.~\ref{einschwing} we show results of measurements for $\varphi(t)$ for the driven pendulum starting at $t=0$ at $\varphi=0$ with some initial velocity. The effect of the transients is clearly visible for $0\leq t\leq 15\,$s. After this time the pendulum oscillates with the frequency of the external torque. 

\subsection{Chaos}
\label{sec:chaos}

\noindent
Perhaps the most interesting phenomenon of a driven pendulum is chaotic motion \cite{Baker1996a}. 
One peculiarity of chaotic motion is that the oscillation of the pendulum will never become periodic, in spite of the fact that it is driven by a periodic force. 
As long as the amplitude of the driving force is small enough, the pendulum behaves like a harmonic oscillator as discussed in Sect.~\ref{sec:regular}. By increasing the amplitude of the driving force the behaviour of the pendulum becomes non-harmonic. In this state it is possible to observe both periodic and chaotic behaviour depending on the values of the other parameters in the equation of motion.

\begin{figure}[tb]
\begin{center}
\includegraphics[width=0.8\columnwidth]{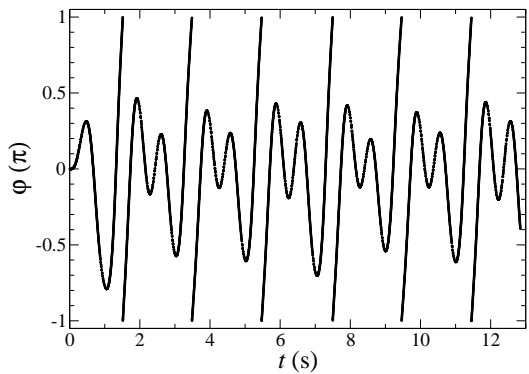}
\caption{\small 
Elongation $\varphi$ as function of the time $t$ for a nonlinear but periodic oscillation of the pendulum.}
\label{2phasen1}
\end{center}
\end{figure}
\begin{figure}[tb]
\begin{center}
\includegraphics[width=0.8\columnwidth]{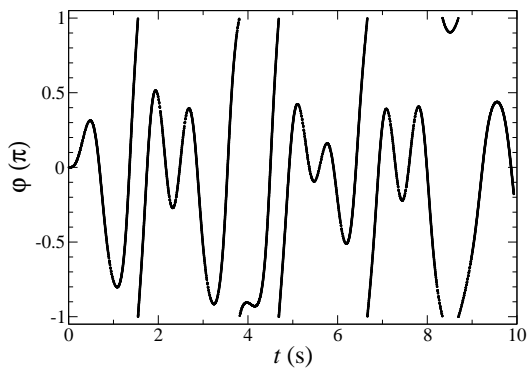}
\caption{\small 
Elongation $\varphi$ as function of the time $t$ for a chaotic motion of the pendulum.}
\label{ch_w_z1}
\end{center}
\end{figure}

Figure~\ref{2phasen1} shows a nonlinear but periodic oscillation, while the motion of the pendulum in Fig.~\ref{ch_w_z1} is chaotic. 
A chaotic oscillation can only be reached if the pendulum turns over the top position at least once. A critical situation is reached when the pendulum just reaches the top with zero velocity there. If the pendulum stops infinitesimally before the top it will fall back to the side where it came from. If the pendulum stops infinitesimally after reaching the top it will continue its motion on the other side. Then, the dependence of the motion of the pendulum on the initial conditions is so strong, that predicting the long term behaviour is practically impossible. This is the regime of deterministic chaos \cite{Baker1996a}. 

\begin{figure}[tb]
\begin{center}
\includegraphics[width=0.8\columnwidth]{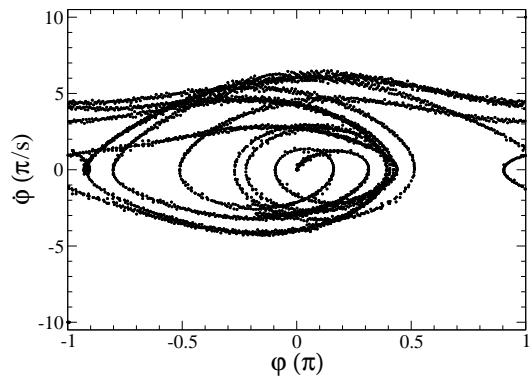}
\caption{\small 
Phase-space representation of a chaotic motion of the pendulum.}
\label{ch_phase1}
\end{center}
\end{figure}

\begin{figure}[tb]
\begin{center}
\includegraphics[width=0.8\columnwidth]{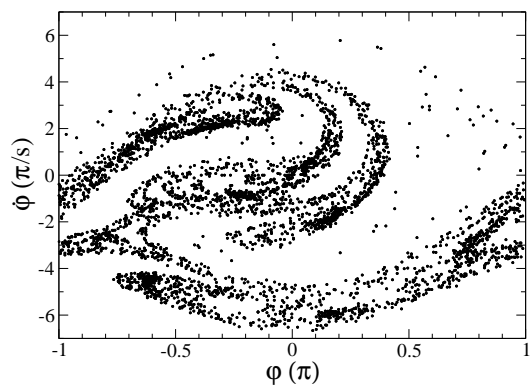}
\caption{\small 
Chaotic motion of the pendulum in phase space represented by a Poincar\'e map.}
\label{chaos}
\end{center}
\end{figure}

To analyse the nonlinear behaviour of a driven pendulum it is useful to study its motion in phase space. This is here the two-dimensional space of position $\varphi$ and canonically conjugate momentum, in our case equal to $\dot\varphi$, the angular velocity. Our device allows us to plot the measured values of the angular velocity of the pendulum vs.~its position for each time step. 
Figure~\ref{ch_phase1} shows the phase-space diagram of a chaotic oscillation. 
The differences in the phase-space representations of periodic and chaotic oscillations are more obvious if the continuous curve is turned into a discrete one. 
This can be done by mapping only the points $\big(\varphi(t),\dot\varphi(t)\big)$ for times $t$ where the phase of the driving torque has some fixed value. This is equivalent to a stroboscopic view of the motion of the pendulum in phase space at the frequency of the driving force. 
The result is called a ``Poincar\'{e} map.'' It is obvious, that the number of points of the Poincar\'{e} map is limited if the motion of the pendulum is periodic. 
Theory predicts that the Poincar\'{e} map of a chaotic oscillation should show fractal characteristics \cite{Baker1996a}. 
Figure~\ref{chaos} shows a Poincar\'{e} map with 4000 data points from a chaotic motion of the pendulum. Our experimental results are in complete agreement with computer simulations made by one of us (T. G.) following Baker and Gollub \cite{Baker1996a}. 

\section{Conclusion}
\label{conclusion}
We have constructed and operated a versatile device for demonstrating the rich range of motional phenomena possible with a damped and driven physical pendulum. 
The degree of damping, restoring, and driving by an external periodic torque, can all be determined by a computer controlled torsion spring. 
A computer readout of the position data allows for detailed offline study of the motion. 
The device can serve as a teaching tool at schools and universities to demonstrate all aspects of oscillatory motion ranging from simple harmonic motion to chaotic behaviour.

\section*{Acknowledgement}
We thank M. Oehmig for contributing to this project in its initial stage.


\end{document}